# Approaches for benchmarking single-cell gene regulatory network inference methods

Yasin Uzun[1]

[1] Penn State College of Medicine, Department of Pediatrics, Department of Biochemistry and Molecular Biology, Hershey, PA, USA

## Abstract

Gene regulatory networks are powerful tools for modeling interactions among genes to regulate their expression for homeostasis and differentiation. Single-cell sequencing offers a unique opportunity to build these networks with high-resolution data. There are many proposed computational methods to build these networks using single-cell data and different approaches are followed to benchmark these methods. In this review, we lay the basic terminology in the field and define the success metrics. Next, we present an overview of approaches for benchmarking computational gene regulatory network approaches for building gene regulatory networks and point out gaps and future directions in this regard.

## Introduction

Gene regulatory networks (GRNs) help to model the interactions that regulate gene expression, thereby providing a holistic view of the genetic landscape. Profiling these interactions is important for understanding the pathways related to developmental processes and various diseases. GRNs also provide topology information regarding individual genes, which is indicative of their molecular roles in the cell.

Building these networks accurately has been an active area of research in systems biology. Next-generation sequencing (NGS) methods provide high-throughput data that enable system-wide construction of these networks by including the entire set of genes, although they can only be constructed using efficient computational methods due to their immense size. The methods employ different approaches, including correlation-, mutual information-, and regression-based techniques, probabilistic methods, and Bayesian models [1–13].

Based on NGS technology, single-cell genomic sequencing methods provide the genetic status of individual cells, generating data with the utmost resolution. The advent of single-cell sequencing methods has also resulted in a burst of methodologies for building GRNs using this type of data [14–37]. These methodologies employ a diverse set of approaches for this purpose, which have been discussed extensively elsewhere [38–44]. However, most previous studies have utilized a different approach for benchmarking their methods, making comparisons across the studies infeasible. Efforts for independent benchmarking studies are highly limited and report unsatisfactory levels of success and reproducibility [45,46], in contrast to those reported in method studies.

The lack of a common understanding of benchmarking GRN methods causes discrepancies in the success rates across different studies and hampers progress in the field. Here, we aim to provide a general overview of common approaches for benchmarking GRNs based on single-cell genomic data. Common approaches for benchmarking single-cell GRNs include using simulated datasets, protein interaction networks, and regulatory databases. Each approach has its advantages and limitations, and can be preferable depending on the type of GRN category that is benchmarked. In the rest of this paper, we set the terminology and discuss common benchmarking approaches.

## Classification of GRNs

Although there are a considerable number of published GRN inference methods, there is a lack of a clear definition for the terminology being used for this field, starting from the GRN term itself. In the context of this paper, we define a GRN as the set of *directed* interactions between any type of gene, and these interactions are represented with edges that originate from the regulator gene and destined to the gene that is being regulated. As the genes are biologically not functional themselves, gene regulation is performed by the product of the regulator gene, which may be proteins or other molecules generated from the genes.

Based on this definition, there are two critical characteristics of GRNs. 1) All the nodes are genes. 2) All the edges are directed. The second characteristic draws a clear boundary between GRNs and another type of commonly used genetic network type: gene co-expression networks (GCNs). Although GCNs also consist of genes, the edges in the GCNs are not directed, as opposed to GRNs. The edges in a GCN represent the co-expression relationships between the two nodes, which represent the genes in the network. Although GCNs are valuable tools for detecting functionally related genes performing different biological processes in the cell, unlike GRNs, they do not provide causality information. GCNs do not clarify whether one of the two connected genes is regulating the other, or whether they are regulated by a third gene [47]. We limit our discussion to benchmarking GRNs in this paper, as their directionality information has a dramatic impact on how to assess the accuracy of a network.

We want to note that GRNs and transcriptional (or transcription-) regulatory networks (TRNs) are often used interchangeably in the literature. For clarification, we define GRNs as networks that model all types of interactions that regulate gene expression. In this context, the edges in a GRN can originate from any gene without distinction. The term TRN, however, will be used for a specific subset of GRNs in which directed edges originate from transcription factor (TF) genes only, and these edges represent the transcriptional regulation relationship. This distinction will help establish a common language and terminology in the field. Based on this definition, most approaches for benchmarking GRNs can also be used for TRNs, as subsetting the originating nodes by selecting TFs only in a directed GRN yields a TRN. However, this can be a questionable approach in terms of efficiency, as this approach ignores the binding specificity of TFs, which is one of their main mechanisms of action.

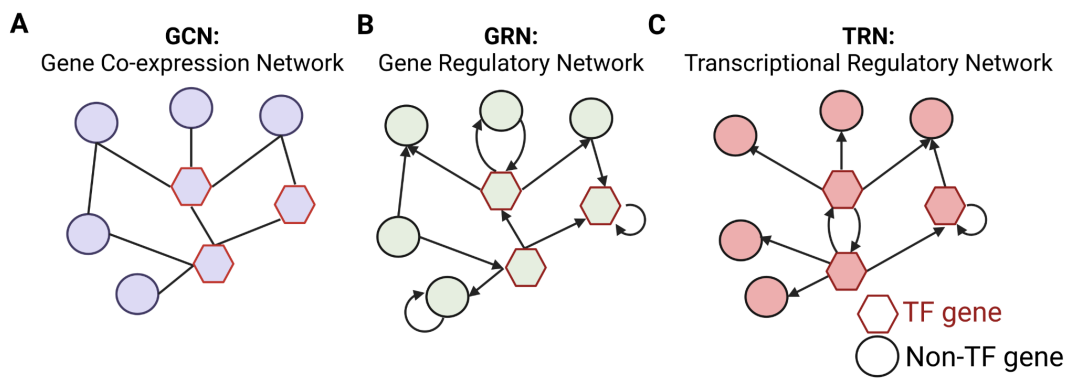

**Figure 1**. **Different types of gene networks. A)** GCN: Network with undirected edges. **B)** GRN: Network with directed edges. **C)** TRN: Network with directed edges that can only originate from transcription factors (TFs).

**Accuracy metrics**

To assess the accuracy of a network mathematically, the network must be represented in a numerical format. This is often represented using an adjacency matrix in which the rows are the regulator genes, the columns represent the regulated genes, and the values in the matrix entries represent the interactions between the genes. Scaling the edge weights in the network between 0 and 1 enables practical comparison and ease of interpretability for the user, as numerical values without limits are hard to compare or difficult to interpret. The directionality information can be further divided into positive (stimulatory) or negative (inhibitory) interactions via the sign of the entry value, although not all network construction methods are able to infer this information. For networks with this type of information, the edge weights can be scaled between -1 and 1.

Assessing the accuracy of a GRN essentially compares the network topology built by a network method to a gold standard benchmark network. The edge weights for a gold standard GRN can take continuous values or alternatively binary (0,1) or ternary (-1,0,1) values, depending on the methodology used to derive these networks. Adjacency matrices of the built and gold standard networks can be compared in multiple ways. One way is to simply compute the error rate by measuring the absolute difference between the two adjacency matrices and computing the sum or average across the matrix difference.

Most approaches for measuring network accuracy are based on comparing the constructed networks with the ground truth networks as the gold standard. When both the tested and the gold standard networks are in a pruned state (with edges having discrete values), the correct and erroneous edge inferences can be quantified by pairwise comparison of edges in terms of true positives (TP), false positives (FP), true negatives (TN), and false negatives (FN) [48]. Using these values, various error metrics can be calculated, including precision, recall (sensitivity), specificity [49], and percentage of interaction (PerInt) [46] (Figure 2A).

It may not always be possible to obtain a pruned network, due to the network construction method lacking a specified threshold for removing edges with low predictive power. Even if it may be technically possible, it may not be preferable, as the recall and precision are adversely correlated; different methods may have one of them better than the other, making comparison of the methods difficult. Precision-recall (PR) [50] and receiver operating characteristic (ROC) curves [51] are viable metrics and offer a solution to this problem by removing the necessity of pruning or thresholds. Both methods compute two different success metrics for varying thresholds for the comparison of a network with continuous edge weights to a ground truth binary network. These two success metrics are depicted as a curve, and the area under the curve (AUC) is used to assess the accuracy of the constructed network. In addition to removing the necessity of pruning and thresholding, these methods also have the advantage of offering a balanced view of accuracy in terms of precision and recall (Figure 2B).

Despite its advantages, computing AUC may not be possible when the gold standard network is not in the pruned (binary) state, specified with edge weights in a continuous spectrum. The weighted Jaccard index (WJI) [46] offers a solution for such cases by comparing the sum of pairwise minimum edge weights to the maximum (Figure 2C). However, this metric is sensitive to the scale of the edge weights, and if the range of the edge weights is considerably larger in one network than in the other, the metric can be biased toward 0. To avoid this pitfall, the edge weights in the two networks need to be scaled into the same range.

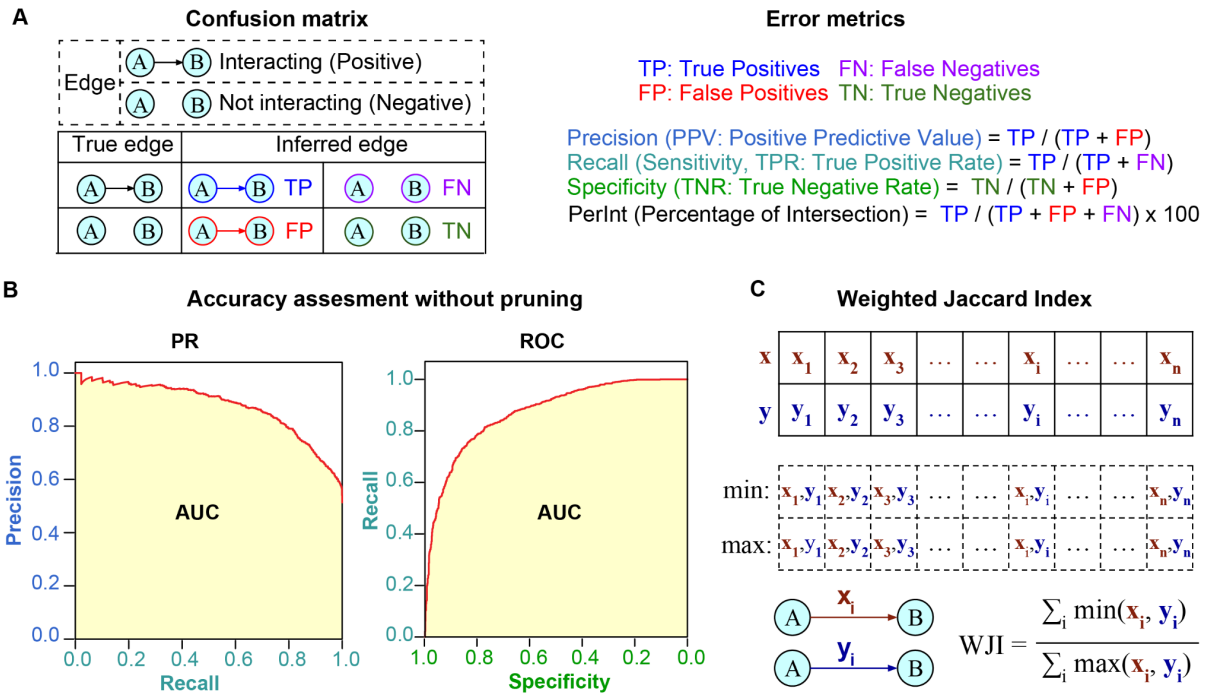

**Figure 2. Methods and metrics for assessing the accuracy of GRNs using a binary gold standard network as the benchmark.**
**A** Computation of the confusion matrix and error metrics when comparing a pruned network (with discrete edge weights) to a gold standard binary network, shown with a hypothetical interaction between the two genes A and B.
**B** Methods for assessing the overall accuracy of an unpruned network (with continuous edge weights) by comparing it to a gold standard binary network. **PR**: precision-recall, **ROC**: receiver operating characteristics, **AUC**: area under the curve.
**C** Weighted Jaccard index (WJI) for comparing two networks with continuous edge weights. $x_i$ and $y_i$ represent the weights of the edges between the gene pairs.

The accuracy metrics described above depend on comparing the constructed networks with a ground-truth dataset. In addition to these metrics, there are indirect approaches to benchmarking GRNs. These include deriving the master regulators from the networks and comparing them with the literature, applying clustering using the network data and benchmarking the clustering [52], comparing the regulons with the known pathways [36] or measuring the connectivity in the network [40]. Although these approaches can also be helpful, they are relatively more challenging to standardize and interpret than ground truth approaches.

### Ground truth

All the accuracy metrics described above rely on a ground truth dataset for benchmarking accuracy. This is a significant challenge in GRN research, as it is highly difficult to construct a robust network even directly through experimental techniques, especially for higher organisms, such as vertebrates. At a minimum, gain and loss of function experiments are required for each individual regulator. To specifically model TRNs, TF bindings also need to be identified via chromatin immunoprecipitation followed by sequencing (ChIP-seq) or an alternative assay. As a result, the current approaches for benchmarking GRNs rely on simple model organisms, such as *Escherichia coli* and yeast. For complex organisms, ground truth networks are usually built for a specific tissue/cell type and a limited set of regulators.

In this context, there are several alternative approaches to building ground truth GRNs (Figure 3). They include 1) generating synthetic networks based on simple model organisms via specialized software, 2) using existing protein–protein interaction networks, and 3) using existing regulatory databases. Each approach has its own advantages and disadvantages, and each can be used in a certain context.

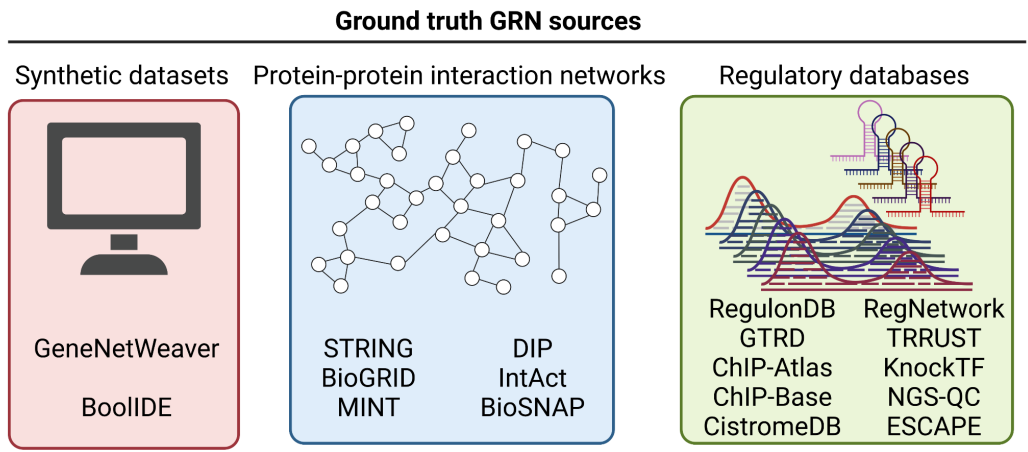

**Figure 3**. **Different approaches for obtaining ground truth GRNs.** Synthetic datasets are generated via specialized software. PPI networks and regulatory databases are other alternative ground truth networks.

## Synthetic datasets

Generating synthetic datasets is widely used for benchmarking computational methods for many different problems. GeneNetWaiver software was introduced to simulate gene regulatory networks and generic synthetic datasets in this context [53]. Used for the DREAM5 challenge to assess the accuracy of GRN inference methods, GeneNetWaiver is built upon topologies from the known interaction networks of *E. coli* [54] and *Saccharomyces cerevisiae* [55]. It uses a thermodynamic approach to model transcript and protein concentration levels, and the noise is modeled through the chemical Langevin equation (CLE) and Gaussian white-noise processes [56].

GeneNetWaiver generates gene expression datasets for the two aforementioned organisms, which are then used to benchmark the network construction method. Although this tool is widely used, its obvious limitation is that the networks of bacteria or yeast are far from ideal for modeling complex organisms, such as mammals, even without considering the tissue specificity of these dynamic networks. Due to the age of the tool, it cannot simulate single-cell omics data, and its applicability is very limited.

An alternative tool for generating synthetic gene regulatory networks is BoolIDE [57]. This simulator models GRNs as a system of stochastic differential equations and generates synthetic single-cell expression data. Both GeneNetWaiver and BoolIDE cannot model epigenetic control and are limited to expression-based control. Hence, although they can be applied to derive general-purpose GRNs, they are not in a position to model TRNs accurately.

**Protein–protein interaction networks**

Gene regulation takes place via the proteins that are coded by the regulating genes, and it can be viewed as the interaction of two proteins, where one protein (TF) activates or inhibits the other (target). In this context, GRNs can be viewed as a subset of protein–protein interaction (PPI) networks and can be compared to PPI networks for assessing their accuracy [45]. There are various protein–protein interaction databases, including STRING [58], Biological General Repository for Interaction Datasets (BioGRID) [59], the Molecular INTeraction database (MINT) [60], the Database of Interacting Proteins (DIP) [61], the IntAct molecular interaction database (IntAct) [62]. and BioSNAP [63] (Table 1).

Although the robustness of PPI networks is a significant advantage, most of these networks are static and do not have tissue specificity. For this reason, most of them can only help to benchmark common GRNs of the entire organisms but not for a specific cell type or tissue. However, future updates to these repositories can introduce filtering by tissue and cell type to address this issue [58].

**Regulatory databases**

The transcriptional regulatory databases provide access to regulatory interactions. These databases provide lists of TF binding sites and targets using peak regions inferred from chromatin data, such as ChIP-Seq, DNase-Seq, ATAC-Seq, FAIRE-Seq, and MNase-Seq. As a result, these databases can be a valuable source for benchmarking TRNs, which are specialized GRNs with edges originating from TFs only. These repositories include RegulonDB [64], Gene Transcription Regulation Database (GTRD) [65], ChIP-Atlas [66], ChIPBase [67], CistromeDB [68], RegNetwork [69], TRRUST [70], KnockTF [71], NGS-QC [72], and ESCAPE [73] (Table 2).

These repositories present lists of TF target genes based on the application of various peak calling methods on curated experiments. This feature allows for modeling of the epigenetic regulation mechanism, which cannot be performed by approaches that are purely based on expression data. This characteristic is particularly important since the epigenetic state of the chromatin and the DNA affects the binding affinity of the TFs to the regulatory regions, and any model that discards this information cannot model the TRNs properly. However, one limitation of these repositories is that, as they are collections of TF binding experiments, their scope is limited to information available in extant studies, as the targets of a specific TF for a certain cell type or tissue are only available if it has been studied experimentally.

The regulatory databases have another important characteristic that limits their power. Although the chromatin state experiments present valuable information about TF binding sites (TFBS) in DNA, they do not provide any information about the specific gene that is being targeted by the TF. To determine the regulating TFs, these databases scan several thousand or tens of thousands of base pairs near the transcription start sites (TSS) of the genes. Although practical to implement, scanning the flanking regions of TSS for TFBS is not an accurate model of epigenetic regulation, as the enhancers to which TFs bind for gene regulation can interact with genes that are far more distant linearly (in terms of base pairs). An alternative approach can be to enlarge the scanning region for TF binding peaks. However, this would result in an ever-increasing number of regulating genes for a target gene in parallel to the size of the region being used, reducing the specificity.

**Discussion**

Methods for building GRNs are heterogeneous due to the specific purpose and domains that these methods are expected to serve. Although the described qualitative criteria can be applied to all methods, the diversity of the methods makes it necessary to apply a tailored approach to benchmarking the accuracy of the different categories of constructed networks.

The choice of the ground truth network is dependent on the organism under study. There are experimentally constructed robust ground truth GRNs for simple organisms, but for complex organisms, this can only be accomplished for a specific cell time with limited accuracy. The specific type of GRN also has implications for the selection of ground truth networks. The datasets derived from loss and gain of function experiments can be a reliable source of information for identifying interactions for general-purpose GRNs. However, specifically for TRNs, additional epigenetic data, such as ChIP-seq, are needed to identify TF binding sites.

At this point in the state of the art, there is a gap for a central repository for ground truth GRNs, considering all the characteristics described in this review. Similar to the contribution of the University of California Irvine Machine Learning Repository [74] to the field of machine learning, a repository for GRNs will serve investigators studying network construction methods in an unbiased manner and advance the field. This repository can also help define a standard file format for generated GRNs, which will help to avoid confusion in interpreting the output of every individual tool in this context. In our view, a plain text format containing the minimal information possible, such as the interacting gene identifiers and numerical edge weights, will facilitate easy interpretation. The guidelines that are represented here can help to design and implement such a GRN repository, which can serve as a useful resource for the field in the long term.

There are various criteria for benchmarking gene regulatory construction methods. Although the accuracy of the network topologies is heavily discussed in most studies, the alternative criteria, including the run-time performance and practical use, are often overlooked for benchmarking different network methodologies, although they may also be reasons for preferring one specific method over another, especially if the reported accuracy rates are close.

The increased resolution of the single-cell datasets also leads to an immense increase in data dimensionality and size. Without scalable computational implementation, even a highly accurate GRN method will have limited use, as the sizes of single-cell datasets are increasing rapidly and may increase exponentially when multiple samples are combined. For this reason, it is important to design GRN construction methods in an efficient manner. In parallel, benchmarking approaches should take processor and memory requirements, as well as computational run times, into account when evaluating different methods.

Although quantifiable assessments, such as accuracy, processor, memory, and run-time requirements, are benchmarked in detail, the qualitative characteristics that define the level of usability are often overlooked during benchmarking. However, these are valuable guidance for users of any computational framework. Even the methods with the best accuracy can be of little benefit unless they have an efficient design that enables practical usage with a limited need for troubleshooting. In this context, ideally, independent benchmarking approaches should assess the following usability characteristics (Figure 4A):

1) Availability: Whether publicly accessible software is available for download.

2) Transparency: Whether the source code for the implementation is available and implemented in an open-source programming language.

4) Documentation: Whether proper documentation describing is available for installation, usage, and replicating of the results reported by the authors.

5) Reproducibility: Whether the raw and processed data are available for reproducing the results reported by the authors.

6) Support: Whether troubleshooting support is provided by the authors.

7) Update: Whether the software is updated to fix reported errors.

In addition to these characteristics, other general principles of scientific software can also be used to assess the usability of the software implementations of GRN methods [75,76]. These qualitative evaluations will guide users in selecting appropriate tools for their needs.

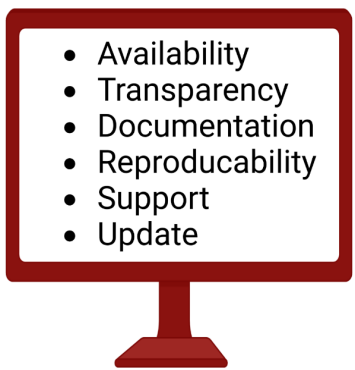
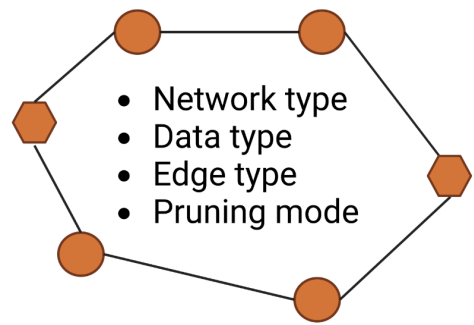

**Figure 4. Qualitative metrics for assessing computational GRN methods.**
**A** General principles for scientific software.
**B** Characteristics specific to the constructed GRNs.

In addition to these qualitative characteristics, benchmarking approaches should assess the following specific characteristics that are specific for GRN inference methods (Figure 4B):

1) GRN type: Whether the method can infer general GRNs or is specialized for TRNs.

2) Data type: Whether the method is based only on gene expression data or utilizes sequence information and epigenetic data as well.

3) Edge type: Whether the edge weights are binary, ternary, or continuous.

4) Pruning mode: Whether the method can prune the network or provide a metric (such as a p-value) for the edges that can be used to prune the network.

Collectively, these criteria will allow interested users to make informed decisions about using a specific tool. We envision that such a comprehensive benchmarking approach is one of the current necessities in this field for facilitating usability. This will further encourage investigators to develop new GRN methods that consider these aspects.

## Conclusion

There has been a considerable amount of research on building GRNs in recent decades. Due to its importance in understanding disease-related pathways and the fast pace of advances in omics technologies, interest in regulatory networks is only expected to increase. Hence, we anticipate that novel methodologies will be proposed for building GRNs. This necessitates establishing a standardized approach to benchmarking these methods.

This review presented a brief overview of the current approaches for benchmarking GRNs, together with their strengths and limitations, and highlighted potential ways of addressing these limitations. The presented knowledge will guide future investigators in establishing benchmarking approaches to assess the accuracy of GRN methods and to develop more accurate and usable tools.

## Declarations

The authors declare that they have no conflict of interest.

**Table 1. Protein interaction databases**

| Database | Organism(s) | URL | PMID |
|---|---|---|---|
| STRING | Human<br>Rat<br>Mouse<br>Zebrafish<br>*D. melanogaster*<br>*Arabidopsis*<br>*C. elegans*<br>*E. coli*<br>*S. cerevisiae*<br>*P. aeruginosa*<br>*Others inferred with orthology | string-db.org | 33237311 |
| BioGRID | Human<br>Rat<br>Mouse<br>Zebrafish<br>*D. melanogaster*<br>*Arabidopsis*<br>*C. elegans*<br>*E. coli*<br>*S. cerevisiae*<br>*71 others | thebiogrid.org | 16381927 |
| MINT | Human<br>Mouse<br>*D. melanogaster*<br>*S. cerevisiae* | mint.bio.uniroma2.it | 11911893 |
| DIP | Human<br>Cow<br>Rat<br>Mouse<br>*D. melanogaster*<br>*Arabidopsis*<br>*C. elegans*<br>*E. coli*<br>*S. cerevisiae*<br>*H. pylori* | dip.doe-mbi.ucla.edu | 10592249 |
| IntAct | Human<br>Rat<br>Mouse<br>Arabidopsis<br>*C. elegans*<br>*E. coli*<br>*S. cerevisiae*<br>*D. melanogaster*<br>*C. jejuni*<br>*Synechocystis* sp.<br>*P. falciparum*<br>*H. pylori*<br>*S. pombe*<br>*B. subtilis*<br>Sars-cov-2 | ebi.ac.uk/intact | 14681455 |
| BioSNAP | Human | snap.stanford.edu/biodata | 25915600 |

**Table 2. Regulatory databases**

| Database | Organism(s) | URL | PMID |
|---|---|---|---|
| RegulonDB | *E. Coli* | regulondb.ccg.unam.mx | 30395280 |
| GTRD | Human<br>Mouse | gtrd.biouml.org | 27924024 |
| ChIP-Atlas | Human<br>Mouse<br>Rat<br>*D. melanogaster*<br>*C. elegans*<br>*S. cerevisiae* | chip-atlas.org | 30413482 |
| ChIPBase | Human<br>Mouse<br>Fly<br>Worm<br>*Arabidopsis* | rna.sysu.edu.cn/chipbase | 36399495 |
| CistromeDB | Human<br>Mouse | dc2.cistrome.org | 30462313 |
| RegNetwork | Human<br>Mouse | regnetworkweb.org | 26424082 |
| TRRUST | Human<br>Mouse | grnpedia.org/trrust | 29087512 |
| KnockTF | Human<br>Mouse | bio.liclab.net/KnockTFv2 | 31598675 |
| NGS-QC | Human<br>Chimpanzee<br>*Gallus gallus*<br>Mouse<br>Rat<br>Zebrafish<br>*D. Melanogaster*<br>*Arabidopsis*<br>*C. elegans*<br>*S. cerevisiae* | ngsqc.org | 24038469 |
| ESCAPE | Mouse | maayanlab.net/ESCAPE | 25122140 |